\documentstyle[preprint,aps,epsfig]{revtex}
\begin{document}
\def\tb{{b}}
\title{ Full transverse-momentum spectra of 
low-mass Drell-Yan  pairs at LHC energies}
\author{George Fai$^a$,  Jianwei Qiu$^b$ and Xiaofei Zhang$^a$ }

\address{$^a$ Center for Nuclear Research, Department of Physics,
Kent State University, Kent, Ohio 44242, USA.\\
$^b$ Department of Physics and Astronomy,
Iowa State University, Ames, Iowa 50011, USA.}
\maketitle

\begin{abstract}
The transverse momentum  distribution 
of low-mass Drell-Yan pairs is calculated  in QCD perturbation theory
 with all-order  resummation.  We argue that at LHC energies 
the results  should be reliable for the entire 
transverse momentum  range. We demonstrate that 
the transverse momentum   distribution of low-mass Drell-Yan pairs 
is an advantageous source of constraints on the gluon distribution
and its nuclear dependence.  
\end{abstract}

\section{Introduction}

Dilepton production in hadronic collisions is an excellent 
laboratory for the investigations of strong 
interaction dynamics. This channel provides an opportunity
 for discovery of quarkonium 
states and a clean process for the study of parton distribution functions 
(PDF).  In the Drell-Yan process, the massive lepton-pair is produced via 
the decay of an intermediate $Z^0$ boson or a 
virtual photon $\gamma^*$ with mass $M$.  
When $M\sim M_Z$, the high mass dilepton production in heavy ion 
collisions at LHC energies is dominated by the $Z^0$ channel and 
is an excellent hard probe of  
QCD dynamics \cite{Zhang:2002yz}.  
In this letter, we demonstrate that the transverse momentum 
distribution of  low-mass ($\Lambda_{\rm QCD}\ll M\ll M_Z$) 
dilepton production at LHC energies is a reliable probe of 
both hard and semihard physics at LHC energies and 
is an advantageous source of constraints on gluon distribution in 
the proton and in nuclei.  
In addition, it provides an important contribution to dilepton spectra 
at the LHC, which is the appropriate 
channel to study $J/\psi$, heavy quarks etc.

\section{Full  $p_T$ spectrum  of  low-mass Drell-Yan   Pairs}
In Drell-Yan production,
if both, the physically measured dilepton mass $M$ and the
transverse momentum $p_T$ are large, the cross section in a collision 
between hadrons (or nuclei) $A$ and $B$,
$A(P_A)+B(P_B)\rightarrow \gamma^*(\rightarrow l\bar{l})+X$, 
can be factorized systematically in QCD perturbation theory and 
expressed as
\cite{Collins:gx}
\begin{equation}
\frac{d\sigma_{AB\rightarrow l\bar{l}(M)X}}{dM^2\, dy\, dp_T^2}
=\left(\frac{\alpha_{em}}{3\pi M^2}\right)
 \sum_{a,b}\int dx_1 \phi_{a/A}(x_1,\mu) 
           \int dx_2 \phi_{b/B}(x_2,\mu)\,
 \frac{d\hat{\sigma}_{ab\rightarrow \gamma^* X}}{dp_T^2\, dy}
 (x_1,x_2,M,p_T,y;\mu) .  
\label{Vph-fac}
\end{equation}
The sum $\sum_{a,b}$ runs over all parton flavors; $\phi_{a/A}$
and $\phi_{b/B}$ are normal parton distributions; and $\mu$ represents
the renormalization and factorization scales.  The 
$d\hat{\sigma}_{ab\rightarrow \gamma^* X}/dp_T^2 dy$ in 
Eq.~(\ref{Vph-fac}) is the short-distance probability for partons of 
flavors $a$ and $b$ to produce a virtual photon of invariant mass $M$
and is calculable perturbatively in terms of a power
series in $\alpha_s(\mu)$.  The scale $\mu$ is 
of the order of the energy exchange in the reaction, 
$\mu \sim \sqrt{M^2 + p_T^2}$.  

The transverse momentum ($p_T$) distribution of the dileptons 
can be divided into three regions: low $p_T$ ($\ll M$), intermediate 
$p_T$ ($\sim M$), and high $p_T$ ($\gg M$) regions.  
When both the physically measured $M$ and $p_T$ are large and are of the 
same order, the short-distance partonic part  
$d\hat{\sigma}_{ab\rightarrow \gamma^* X}/dp_T^2 dy$ in 
Eq.~(\ref{Vph-fac}) can be calculated reliably in conventional
fixed-order QCD perturbation theory in terms of a power series in
$\alpha_s(\mu)$.  
However, when $p_T$ is very different from $M$, 
the calculation of Drell-Yan production in both low and high $p_T$ 
regions becomes a two-scale problem in perturbative QCD, and 
the calculated partonic parts include potentially 
large logarithmic terms proportional to a 
power of $\ln(M/p_T)$.  As a result, the higher order corrections in powers 
of $\alpha_{s}$ are not necessary small.  
The ratio $\sigma^{NLO}/\sigma^{LO}$ [$\propto \alpha_s \times$
(large logarithms)] can be of order 1, 
and convergence of the conventional perturbative expansion 
in powers of $\alpha_{s}$ is possibly impaired.  

In the low $p_T$ region, there are two powers of $\ln(M^2/p_T^2)$ 
for each additional power of $\alpha_s$,
and the Drell-Yan $p_T$ distribution calculated in fixed-order QCD 
perturbation theory is known not to be reliable.  Only after all-order
resummation of the large $\alpha_s^n\,\ln^{2n+1}(M^2/p_T^2)$ 
do predictions for the $p_T$ 
distributions become consistent with data 
\cite{Qiu:2000hf,Landry:1999an}. 
We demonstrate in Sec.~III that low-mass Drell-Yan production at 
$p_{T}$ as low as $\Lambda_{\rm QCD}$ at LHC energies can be 
calculated reliably in perturbative QCD with all order resummation.

When $p_T \ge M/2$, the lowest-order virtual photon
``Compton'' subprocess: $g+q \rightarrow \gamma^* + q$ dominates the 
$p_{T}$ distribution, and the high-order contributions including
all-order resummation of $\alpha_s^n\, \ln^{n-1}(M^2/p_T^2)$ preserve the fact
that the $p_{T}$ distributions of low-mass Drell-Yan  pairs 
are dominated by gluon initiated partonic subprocesses 
\cite{Berger:2001wr}.  We show in Sec.~IV that the $p_{T}$ distribution of 
low-mass Drell-Yan pairs can be a good probe of the gluon distribution 
and its nuclear dependence.  We give our conclusions in Sec.~V.

\section{Low transverse  momentum   region}

Resummation of large logarithmic terms at low $p_T$ can be carried 
out in either $p_T$ or impact parameter ($\tilde{b}$) space, 
which is the Fourier conjugate of $p_T$ space.  All else being 
equal, the $\tilde{b}$ space approach has the advantage that transverse
momentum conservation is explicit.  Using renormalization
group techniques, Collins, Soper, and Sterman (CSS)~\cite{Collins:1984kg}
devised a $\tilde{b}$ space resummation formalism that resums all 
logarithmic terms as singular as $(1/p_T^2)\ln^m(M^2/p_T^2)$ 
when $p_T \rightarrow 0$.  This formalism has been used widely for 
computations of the transverse momentum distributions of vector bosons 
in hadron reactions \cite{Berger:2002ut}.

At low-mass $M$, Drell-Yan transverse momentum distributions
calculated in the CSS $\tilde{b}$-space resummation formalism  
strongly depend on the non-perturbative parameters at fixed target
energies.   However, it was pointed out recently that the predictive
power of perturbative QCD (pQCD) resummation improves with total
center-of-mass energy $(\sqrt{s})$, and when the energy is high enough,
pQCD should have good predictive power even for low-mass Drell-Yan
production \cite{Qiu:2000hf}. The LHC will provide us a chance to 
study low-mass Drell-Yan production at unprecedented energies.

In the CSS resummation formalism, the differential cross section 
for Drell-Yan production in Eq.~(\ref{Vph-fac}) is reorganized 
as the sum
\begin{equation}
\frac{d\sigma_{AB\rightarrow l\bar{l}(M)X}}
     {dM^2\, dy\, dp_T^2}
=
\frac{d\sigma_{AB\rightarrow l\bar{l}(M)X}^{\rm (resum)}}
     {dM^2\, dy\, dp_T^2}
+
\frac{d\sigma_{AB\rightarrow l\bar{l}(M)X}^{\rm (Y)}}
{dM^2\, dy\, dp_T^2}\, .
\label{css-gen}
\end{equation}
The all-orders resummed term is a Fourier transform from 
the $\tb$-space,
\begin{equation} 
\frac{d\sigma_{AB\rightarrow l\bar{l}(M)X}^{\rm (resum)}}
     {dM^2\, dy\, dp_T^2}
= \frac{1}{(2\pi)^2}\int d^2\tb\, 
e^{i\vec{p}_T\cdot \vec{\tb}}\, W(\tb,M,x_A,x_B)
= \frac{1}{2\pi}\int d\tb\, J_0(p_T \tb)\, \tb\, W(\tb,M,x_A,x_B) \,\, ,
\label{css-resum}
\end{equation}
where $J_0$ is a Bessel function, $x_A= e^y\, M/\sqrt{s}$ and 
$x_B= e^{-y}\, M/\sqrt{s}$, with rapidity $y$ and collision energy 
$\sqrt{s}$.  In Eq.~(\ref{css-gen}), 
the $\sigma^{\rm (resum)}$ term dominates the $p_T$ distributions when
$p_T\ll M$, and the $\sigma^{(Y)}$ term gives corrections that are
negligible for small $p_T$, but become important when $p_T\sim M$.

The function $W(\tb,M,x_A,x_B)$ resums to all orders in QCD perturbation
theory the singular terms that would otherwise behave as $\delta^2(p_T)$ 
and $(1/p_T^2)\ln^m(M^2/p_T^2)$ in transverse momentum space, 
for all $m\ge 0$, and can be calculated perturbatively 
for small \ $\tb$ , 
\begin{equation}
W(\tb,M,x_A,x_B) =
{\rm e}^{-S(\tb,M)}\, W(\tb,c/\tb,x_A,x_B) 
\equiv W^{\rm pert}(\tb,M,x_A,x_B)
\,\,\, ,
\label{css-W-sol}
\end{equation}
where all large logarithms from $\ln(c^2/\tb^2)$ to $\ln(M^2)$ have
been completely resummed into the exponential factor
$S(\tb,M)=\int_{c^2/\tb^2}^{M^2} d\mu^2/\mu^2\left[\ln(M^2/\mu^2)
{\cal A}(\alpha_s(\mu)) + {\cal B}(\alpha_s(\mu))\right]$ with 
functions $\cal A$ and $\cal B$ 
given in Ref.~\cite{Collins:1984kg}, and
$c=2e^{-\gamma_E}$ with Euler's constant $\gamma_E\approx 0.577$.
The function $W(\tb,c/\tb,x_A,x_B)$ in 
Eq.~(\ref{css-W-sol}) is given in terms of modified parton 
distributions from hadron $A$ and $B$ \cite{Collins:1984kg}.  
With only one large momentum scale $1/\tb$, $W(\tb,c/\tb,x_A,x_B)$
is perturbatively calculable.  Since the perturbatively 
resummed $W^{\rm pert}(\tb,M,x_A,x_B)$ in
Eq.~(\ref{css-W-sol}) is only reliable for the small $\tb$ region, an
extrapolation to the nonperturbative large $\tb$ region is necessary 
in order to complete the Fourier transform in Eq.~(\ref{css-resum}).  

In the original CSS formalism, a variable $\tb_*$
and a nonperturbative function $F_{CSS}^{NP}(\tb,M,x_A,x_B)$
were introduced to extrapolate the perturbatively calculated
$W^{\rm pert}$ into the large $\tb$ region such that the full
$\tb$-space distribution was of the form 
\begin{equation}
W^{\rm CSS}(\tb,M,x_A,x_B) \equiv
W^{\rm pert}(\tb_*,M,x_A,x_B)\,
F_{CSS}^{NP}(\tb,M,x_A,x_B)\, ,
\label{css-W-b}
\end{equation}
where $\tb_*=\tb/\sqrt{1+(\tb/\tb_{max})^2}$, with 
$\tb_{max} = 0.5$ GeV$^{-1}$.  This construction ensures that 
$\tb_* \leq \tb_{max}$ for all values of $\tb$.

In terms of the $\tb_*$ formalism, a number of functional forms for
the $F_{CSS}^{NP}$ have been proposed.  A simple Gaussian form in 
$\tb$ was first proposed by Davies, Webber, and Stirling (DWS) 
\cite{Davies:1984sp},
\begin{equation}
F_{DWS}^{NP}(\tb,M,x_A,x_B)
=\exp\left\{-(g_1+g_2\ln(M/2M_0))\tb^2\right\},
\label{DS}
\end{equation}
with the parameters $M_0=2$~GeV, $g_1=0.15$~GeV$^2$, and 
$g_2=0.4$~GeV$^2$.
In order to take into account the appearent dependence on collision 
energies, Ladinsky and Yuan (LY) introduced a new functional form 
\cite{Ladinsky:1993zn}, 
\begin{equation}
F_{LY}^{NP}(\tb,M,x_A,x_B)
=\exp\left\{-(g_1+g_2\ln(M/2M_0))\tb^2-g_1\, g_3
\ln(100x_Ax_B)\tb\right\},
\label{LY}
\end{equation}
with $M_0=1.6$~GeV, $g_1=0.11^{+0.04}_{-0.03}$~GeV$^2$, 
$g_2=0.58^{+0.1}_{-0.2}$~GeV$^2$, and $g_3=-1.5^{+0.1}_{-0.1}$~GeV$^{-1}$.  
Recently, Landry, Brook, Nadolsky, and Yuan proposed a modified
Gaussian form \cite{Landry:2002ix},
\begin{equation}
F_{BLNY}^{NP}(\tb,M,x_A,x_B)
=\exp\left\{-\left[g_1+g_2\ln(M/2M_0)+g_1\, g_3 \ln(100x_Ax_B)
             \right]\tb^2\right\},
\label{BLNY}
\end{equation}
with $M_0=1.6$~GeV, $g_1=0.21^{+0.01}_{-0.01}$~GeV$^2$, 
$g_2=0.68^{+0.01}_{-0.02}$~GeV$^2$, and 
$g_3=-0.6^{+0.05}_{-0.04}$.  
All these parameters were obtained by fitting low energy Drell-Yan and 
high energy $W$ and $Z$ data.  Note,  however that
the $\tb_*$ formalism introduces a modification to the perturbative 
calculation, and the size of the modifications strongly depends on 
the non-perturbative parameters in $F^{NP}(\tb,M,x_A,x_B)$, the 
intermediate boson mass $M$, and collision energy $\sqrt{s}$ 
\cite{Berger:2002ut}.  

A remarkable feature of the $\tb$-space resummation formalism is
that the resummed exponential factor $\exp[-S(\tb,M)]$
suppresses the $\tb$-integral when $\tb$ is larger than $1/M$. 
It can be shown using the saddle point method that, for a large
enough $M$, QCD perturbation theory is valid even at $p_T=0$
\cite{Collins:1984kg}.  For high energy heavy boson ($W$, $Z$, and 
Higgs) production, the integrand of $\tb$-integration in 
Eq.~(\ref{css-resum}) at $p_T=0$ is proportional to 
$\tb W(\tb,Q,x_A,x_B)$, which has a saddle point $\tb_{\rm sp}$ 
well within the perturbative region ($\tb_{\rm sp}<\tb_{max}$), 
and therefore, the $\tb$-integration in Eq.~(\ref{css-resum}) is 
dominated by the perturbatively resummed calculation.  The 
uncertainties from the large-$\tb$ region have very little effect
on the calculated $p_T$ distributions, and the resummation formalism 
is of a good predictive power.

On the other hand, in the low energy Drell-Yan production, 
there is no saddle point in the perturbative region for 
the integrand in Eq. (\ref{css-resum}), 
and therefore the dependence of the 
final result on the non-perturbative input is strong 
\cite{Qiu:2000hf,Landry:1999an}.
However, as discussed in Ref.s 
\cite{Qiu:2000hf,Berger:2002ut,Zhang:2002qc}, the value of the 
saddle point strongly depends on the collision energy $\sqrt{s}$, 
in addition to its well-known $M^2$ dependence.  

Figure 1 shows the integrand of the $\tb$-integration in 
Eq.~(\ref{css-resum}) at $p_T=0$ for production of Drell-Yan pairs 
of mass $M=5$~GeV in proton-proton collisions at $\sqrt{s}=5.5$~TeV 
and $b_{max}=0.5$~GeV$^{-1}$.  Different curves represent different 
extrapolations to the large-$\tb$ region.  The three curves (dashed, 
dotted, and dot-dashed) are evaluated using the $\tb_*$ formalism 
with the DWS, LY, and BLNY nonperturbative functions, respectively.
Although these three nonperturbative functions give similar 
$\tb$-space distributions for heavy boson production at Tevatron 
energies, they predict very different $\tb$-space distributions 
for low-mass Drell-Yan production at LHC energies even within
the perturbative small-$\tb$ region.  Since the $\tb$-distribution
in Fig.~1 completely determines the resummed $p_T$ distribution through 
the $\tb$-integration weighted by the Bessel function $J_0(p_T \tb)$,
we need to be concerned with  the uncertainties of the resummed low-mass $p_T$ 
distributions calculated with different nonperturbative functions. 
We use the CTEQ5M parton distribution function\cite{cteq5} throughout.

In order to improve the situation, a new formalism 
of extrapolation (QZ) was proposed \cite{Qiu:2000hf},
\begin{equation}
W(\tb,M,x_A,x_B) = \left\{
\begin{array}{ll}
W^{\rm pert}(\tb,M,x_A,x_B) & \quad \mbox{$\tb\leq \tb_{max}$} \\
W^{\rm pert}(\tb_{max},M,x_A,x_B)\,
F^{NP}(\tb,M,x_A,x_B;\tb_{max},\alpha)
& \quad \mbox{$\tb > \tb_{max}$}
\end{array} \right. \,\, ,
\label{qz-W-sol-m}
\end{equation}
where the nonperturbative function $F^{NP}$ is given by
\begin{eqnarray}
F^{NP}
=\exp\left\{ -\ln(M^2 \tb_{max}^2/c^2) 
\left[ g_1 \left( (\tb^2)^\alpha - (\tb_{max}^2)^\alpha\right) \right.
 \left.   +g_2 \left(\tb^2 - \tb_{max}^2\right) \right]
-\bar{g}_2 \left(\tb^2 - \tb_{max}^2\right) \right\}.
\label{qz-fnp-m}
\end{eqnarray}
Here, $\tb_{max}$ is a parameter to separate the perturbatively 
calculated part from the non-perturbative input, and its role is 
similar to the $\tb_{max}$ in the $\tb_*$ formalism.  The term 
proportional to $g_1$ in Eq.~(\ref{qz-fnp-m}) represents a direct 
extrapolation of the resummed leading power contribution to the large
$\tb$ region; and the parameters $g_1$  and $\alpha$ 
are determined by the continuity of the function  $W(\tb,M,x_A,x_B)$ 
at $\tb_{max}$.  On the other hand, the values of $g_2$ and 
${\bar g}_2$ represent the size of nonperturbative power corrections. 
Therefore, sensitivity on the $g_2$ and ${\bar g}_2$ in this formalism 
clearly indicates the precision of the calculated $p_T$ distributions.

The solid line in Fig.~1 is the result of the QZ parameterization with 
$\tb_{max}=0.5$~GeV$^{-1}$ and $g_2={\bar g}_2=0$.  Unlike in the 
$\tb_*$ formalism, the solid line represents the full perturbative 
calculation and is independent of the nonperturbative parameters for 
$\tb < \tb_{max}$.  The difference between the solid line and the other 
curves in the small $\tb$ region, which can be as large as 40\%, indicates 
the uncertainties introduced by the $\tb_*$ formalism.  

It is clear from the solid line in Fig.~1 that 
there is a saddle point in the perturbative 
region even for the dilepton mass as low as $M=5$~GeV in Drell-Yan 
production at $\sqrt{s}=5.5$ TeV.  At that energy, $x_A, x_B \sim 0.0045$.  
For such small values of $x$, the PDFs have very strong scaling 
violation, which leads to a large parton shower.  It is the large 
parton shower at the small $x$ that strongly suppresses the function  
$W(\tb,c/\tb,x_A,x_B)$ in Eq.~(\ref{css-W-sol}) as $\tb$ increases.
Therefore, for a $\tb_{max}\sim$(a few GeV)$^{-1}$, the  predictive power
of the $\tb$-space resummation formalism depends on the relative size
of contributions from the small-$\tb$ ($\tb<\tb_{max}$) and 
large-$\tb$ ($\tb>\tb_{max}$) regions of the $\tb$-integration in
Eq.~(\ref{css-resum}).  With a narrow $\tb$ distribution peaked within 
the perturbative region for the integrand, the $\tb$-integration in 
Eq.~(\ref{css-W-sol}) is dominated by the small-$\tb$ region, and 
therefore, we expect pQCD to have good predictive power even for low 
$M$ Drell-Yan production at LHC energies.  

Figure~2 presents our prediction to the fully differential cross 
section $d\sigma/dM^2 dy dp_T^2$ for Drell-Yan production 
in $pp$ collisions at $\sqrt{s}=5.5$ TeV.
Three curves represent the different order of contributions in 
$\alpha_s$ to the perturbatively calculated functions 
${\cal A}(\alpha_s)$, ${\cal B}(\alpha_s)$, 
and ${\cal C}(\alpha_s)$ in the resummation formalism
\cite{Giele:2002hx}.
The solid line represents a next-to-next-to-leading-logarithmic (NNLL) 
accuracy corresponding to keeping the functions, 
${\cal A}(\alpha_s)$, ${\cal B}(\alpha_s)$, 
and ${\cal C}(\alpha_s)$ to the order of $\alpha_s^3$, 
$\alpha_s^2$, and $\alpha_s^1$, respectively.  
The dashed line has a next-to-leading-logarithmic (NLL) accuracy with 
the functions, ${\cal A}(\alpha_s)$, ${\cal B}(\alpha_s)$, 
and ${\cal C}(\alpha_s)$ at
the $\alpha_s^2$, $\alpha_s$, and $\alpha_s^0$, respectively, while
the dotted line has the lowest leading-logarithmic (LL) accuracy 
with the functions, ${\cal A}(\alpha_s)$, ${\cal B}(\alpha_s)$, 
and ${\cal C}(\alpha_s)$ at
the $\alpha_s$, $\alpha_s^0$, and $\alpha_s^0$, respectively.
Similar to what was seen in the fixed order calculation, the
resummed $p_T$ ditribution has a $K$-factor about 1.4-1.6 around the peak 
due to the inclusion of the coefficient ${\cal C}^{(1)}$.

In Eq.~(\ref{qz-fnp-m}), in addition to the $g_1$ term from 
the leading power contribution of soft gluon showers, 
the $g_2$ term corresponds to the first power correction from 
soft gluon showers and the $\bar g_2 $ term is from the intrinsic
transverse momentum of the incident parton.  The numerical values of 
$g_2$ and $\bar g_2$ have to be obtained by fitting the data.  
From the fitting of low energy Drell-Yan data and  heavy gauge 
boson data at the Tevatron, we found that the intrinsic transverse 
momentum term dominates the power corrections and 
it has a weak energy dependence.  For  convenience,
we combine the parameters of the $\tb^2$ term as
$G_2= g_2\ln({M^2 \tb_{max}^2/{c^2}}) + \bar{g}_2$.
For $M=5$ GeV and $y=0$, we use $G_2\sim 0.25$ in the discussion
here \cite{Qiu:2000hf}. 
To test the $G_2$ dependence of our calculation, we define 
\begin{equation}
R_{G_2}(p_T) \equiv \left.
\frac{d\sigma^{(G_2)}_{AB\rightarrow l\bar{l}(M)X}(p_T)}
     {dM^2\, dy\, dp_T^2} \right/
\frac{d\sigma_{AB\rightarrow l\bar{l}(M)X}(p_T)}
     {dM^2\, dy\, dp_T^2} \,\,\, ,
\label{Sigma-g2}
\end{equation}
where the numerator represents the result with finite $G_2$,
and the denominator contains no power corrections ($G_2 = 0$).

The result for $R_{G_2}$ is shown in Fig.~3. $R_{G_2}$
deviates from unity less than $1\%$. The dependence of our result on 
the non-perturbative input is indeed very weak. 

Since the $G_2$ terms represent the power corrections from soft gluon
showers and partons' intrinsic transverse momentum, the smallness of 
the deviation of $R_{G_2}$ from 
unity also means that leading 
power contributions from gluon showers dominate the dynamics of 
low-mass Drell-Yan production at LHC energies. 
Even though the power corrections
will be enhanced in nuclear collisions, we expect 
it to be  still less than several percent \cite{Zhang:2002qc}.  
The isospin effects are  also small here, because 
$x_A$ and $x_B$ are  small.

Since the leading power contributions from initial-state parton showers 
dominate the production dynamics, the important nuclear effect is 
the modification of parton distributions.  Because the 
$x_A$ and $x_B$ are small for low-mass Drell-Yan 
production at LHC energies, shadowing is the only dominant 
nuclear effect.  In order to study the shadowing effects, we define
\cite{Zhang:2002qc}
\begin{equation}
R_{sh}(p_T) \equiv \left.
\frac{d\sigma^{(sh)}_{AB\rightarrow l\bar{l}(M)X}(p_T,Z_A/A,Z_B/B)}
     {dM^2\, dy\, dp_T^2} \right/
\frac{d\sigma_{AB\rightarrow l\bar{l}(M)X}(p_T)}
     {dM^2\, dy\, dp_T^2} \,\,\, .
\label{Sigma-sh}
\end{equation}
We plot in Fig.~\ref{fig4} the ratio $R_{sh}$ as a function of
$p_T$ in  $pPb$ and $PbPb$ collisions at $
\sqrt s=5.5$~TeV for $M=5$~GeV and
$y=0$.  The EKS parameterizations of nuclear parton distributions 
\cite{Eskola:1998iy} were used to evaluate the cross sections
in Eq.~(\ref{Sigma-sh}). 
  Fig.~\ref{fig4} shows that 
$R_{sh}$ decreases about 30\% from $pPb$ to $PbPb$ collisions.
It is clear that low-mass Drell-Yan production at  $pPb$ and $PbPb$
can be a good probe of nuclear shadowing. 

\section{High  transverse  momentum region}

The gluon distribution plays a central role in calculating many 
important signatures at hadron colliders because of the dominance of gluon
initiated subprocesses.  A precise knowledge of the gluon distribution
as well as its nuclear dependence is absolutely vital for understanding
both hard and semihard probes at LHC energies.  

It was pointed out recently that the transverse momentum distribution 
of massive lepton pairs produced in hadronic collisions is an 
advantageous source of constraints on the gluon distribution
\cite{Berger:1998ev}, free from the experimental and 
theoretical complications of photon isolation that beset studies of 
prompt photon production~\cite{Berger:1990es,Berger:1995cc}.
Other than the difference between a virtual and a real photon, 
the Drell-Yan process and prompt photon production 
share the same partonic subprocesses.  Similar to prompt photon production, 
the lowest-order virtual photon ``Compton'' subprocess: 
$g+q\rightarrow \gamma^*+q$ dominates the $p_T$ distribution 
when $p_T > M/2$, and the next-to-leading order contributions 
preserve the fact that the $p_T$ distributions are dominated by 
gluon initiated partonic subprocesses~\cite{Berger:1998ev}. 

There is a phase space penalty associated with the finite mass of 
the virtual photon, and the Drell-Yan factor 
$\alpha_{em}/(3\pi M^2)< 10^{-3}/M^2$ in Eq.~(\ref{Vph-fac}) 
renders the production rates for massive lepton pairs small 
at large values of $M$ and $p_T$.  In order to enhance the
Drell-Yan cross section while keeping the dominance of the gluon
initiated subprocesses, it is useful to study lepton pairs with
low invariant mass and relatively large transverse momentum
\cite{Berger:2001wr}.  
With the large transverse momentum $p_T$ setting the hard scale of the 
collision, the invariant mass of the virtual photon $M$ can be small,
as long as the process can be identified experimentally, 
and the numerical value $M\gg\Lambda_{\rm QCD}$.  
For example, the cross section for Drell-Yan
production was measured by the CERN UA1 Collaboration~\cite{Albajar:1988iq}
for virtual photon mass $M\in [2m_\mu, 2.5]$~GeV.  

When $p_T^2\gg M^2$, the perturbatively calculated short-distance
partonic parts, $d\hat{\sigma}_{ab\rightarrow \gamma^* X}/dp_T^2 dy$
in Eq.~(\ref{Vph-fac}), receive one power of the logarithm
$\ln(p_T^2/M^2)$ at every order of $\alpha_s$ beyond the leading
order.  At sufficiently large $p_T$, the coefficients of
the perturbative expansion in $\alpha_s$ will have large 
logarithmic terms, and these high order corrections may not be small.  
In order to derive reliable QCD predictions, resummation of the 
logarithmic terms $\ln^m(p_T^2/M^2)$ must be considered.  It was 
recently shown \cite{Berger:2001wr} that the large
$\ln^m(p_T^2/M^2)$ terms in low-mass Drell-Yan cross sections
can be systematically resummed into 
a set of perturbatively calculable 
virtual photon fragmentation functions \cite{Qiu:2001nr}, and 
similar to Eq.~(\ref{css-gen}), the differential cross section
for low-mass Drell-Yan production at large $p_T$ 
can be reorganized as 
\begin{equation}
\frac{d\sigma_{AB\rightarrow l\bar{l}(M)X}}
     {dM^2\, dy\, dp_T^2}
=
\frac{d\sigma_{AB\rightarrow l\bar{l}(M)X}^{\rm (resum)}}
     {dM^2\, dy\, dp_T^2}
+
\frac{d\sigma_{AB\rightarrow l\bar{l}(M)X}^{\rm (Dir)}}
     {dM^2\, dy\, dp_T^2}\, ,
\label{bqz-gen}
\end{equation}
where $\sigma^{\rm (resum)}$ includes the large logarithms 
and can be factorized as \cite{Berger:2001wr}
\begin{eqnarray}
\frac{d\sigma_{AB\rightarrow l\bar{l}(M)X}^{\rm (resum)}}
     {dM^2\, dy\, dp_T^2}
&=&\left(\frac{\alpha_{em}}{3\pi M^2}\right)
\sum_{a,b,c}
\int dx_1 \phi_{a/A}(x_1,\mu)\, 
\int dx_2 \phi_{b/B}(x_2,\mu)\,
\nonumber \\
&\ & \times
\int \frac{dz}{z^2}\,
\frac{d\hat{\sigma}_{ab\rightarrow c X}}
     {dp_{c_T}^2\,dy}(p_{c_T}=p_T/z)\
  D_{c\rightarrow \gamma^* X}(z,\mu_F^2;M^2) \,\,  ,
\label{DY-F-fac}
\end{eqnarray}
with the factorization scale $\mu$ and fragmentation scale $\mu_F$,
and the virtual photon fragmentation functions
$D_{c\rightarrow \gamma^*}(z,\mu_F^2;Q^2)$.
The $\sigma^{\rm (Dir)}$ term plays the same role as 
$\sigma^{(Y)}$ term in Eq.~(\ref{css-gen}), and it dominates the cross section 
when $p_T\rightarrow M$.

Figure~\ref{fig5} presents the fully resummed transverse momentum spectra
of low-mass Drell-Yan production in $pp$ collisions with $M=5$~GeV 
at $y=0$ and $\sqrt{s}=5.5$~TeV (solid).  For comparison, we also plotted
the leading order spectra calculated in  conventional fixed order pQCD.
The fully resummed distribution is larger in the large $p_T$ region and 
smoothly convergent as $p_T\rightarrow 0$.  In addition, as discussed 
in Ref.~\cite{Berger:2001wr}, the
resummed differential cross section is much less sensitive to the 
changes of renormalization, factorization, and fragmentation scales, 
and should be more reliable than the fixed order calculations.

To demonstrate the relative size of gluon initiated contributions, we 
define the ratio
\begin{equation}
R_g = \left.
\frac{d{\sigma}_{AB\rightarrow \gamma^*(M) X}(\mbox{gluon-initiated})}
     {dp_T^2\, dy} \right/
\frac{d{\sigma}_{AB\rightarrow \gamma^*(M) X}}{dp_T^2\,dy}\, .
\label{Rg}
\end{equation}
The numerator includes the contributions from all partonic subprocesses 
with at least one initial-state gluon, and 
the denominator includes all subprocesses.

In Fig.~\ref{fig6}, we show $R_g$ as a function of $p_T$ in $pp$ 
collisions at $y=0$ and $\sqrt{s}=5.5$~TeV with $M=5$~GeV.  
It is clear from Fig.~\ref{fig6} that gluon 
initiated subprocesses dominate the low-mass Drell-Yan cross section and 
that low-mass Drell-Yan lepton-pair production at large transverse 
momentum is an excellent source of information on the gluon 
distribution~\cite{Berger:2001wr}.
The slow falloff of $R_g$ at large $p_T$ is related to the 
reduction of phase space and the fact that cross sections are 
evaluated at larger values of the partons' momentum fractions.

\section{Conclusions}

In summary, we present the fully differential cross section of low-mass 
Drell-Yan production calculated in QCD perturbation theory 
with all-order resummation.  For $p_T\ll M$, we use CSS $\tb$-space
resummation formalism to resum the large logarithmic contributions
as singular as $ln^m(M^2/p_T^2)/{p_T^2} $
to all orders in $\alpha_s$. We show that the resummed $p_T$ distribution
of low-mass Drell-Yan pairs at LHC energies is 
dominated by the perturbatively calculable small $\tb$-region and thus 
reliable for $p_T$ as small as $\Lambda_{\rm QCD}$.  Because of the 
dominance of small $x$ PDFs, the low-mass Drell-Yan 
cross section is a good probe of the nuclear dependence of 
parton distributions.  For $p_T\gg M$, we use a newly derived 
QCD factorization formalism \cite{Berger:2001wr} to resum 
all orders of $\ln^m(p_T^2/M^2)$ type logarithms.  We show that almost
90\% of the low-mass Drell-Yan cross sections at LHC energies is 
from gluon initiated partonic subprocesses.  Therefore, the low-mass 
Drell-Yan cross section at $p_T>M$ is an advantageous source of information
on the gluon distribution and its nuclear dependence --- shadowing.  
Unlike other probes of gluon distributions, low-mass Drell-Yan 
does not have the problem of isolation cuts associated with direct photon 
production at collider energies, and does not have the hadronization 
uncertainties of $J/\psi$ and charm production. Moreover, 
the precise information on dilepton production from the
Drell-Yan channel is critical for studying charm production at 
LHC energies.

\vspace{0.1in}

\section*{Acknowledgments}

This work was supported in part by the U.S. Department of Energy under 
Grant No.s DE-FG02-86ER-40251 and DE-FG02-87ER-40371.



\begin{figure}
\centerline{\includegraphics[width=6.5cm]{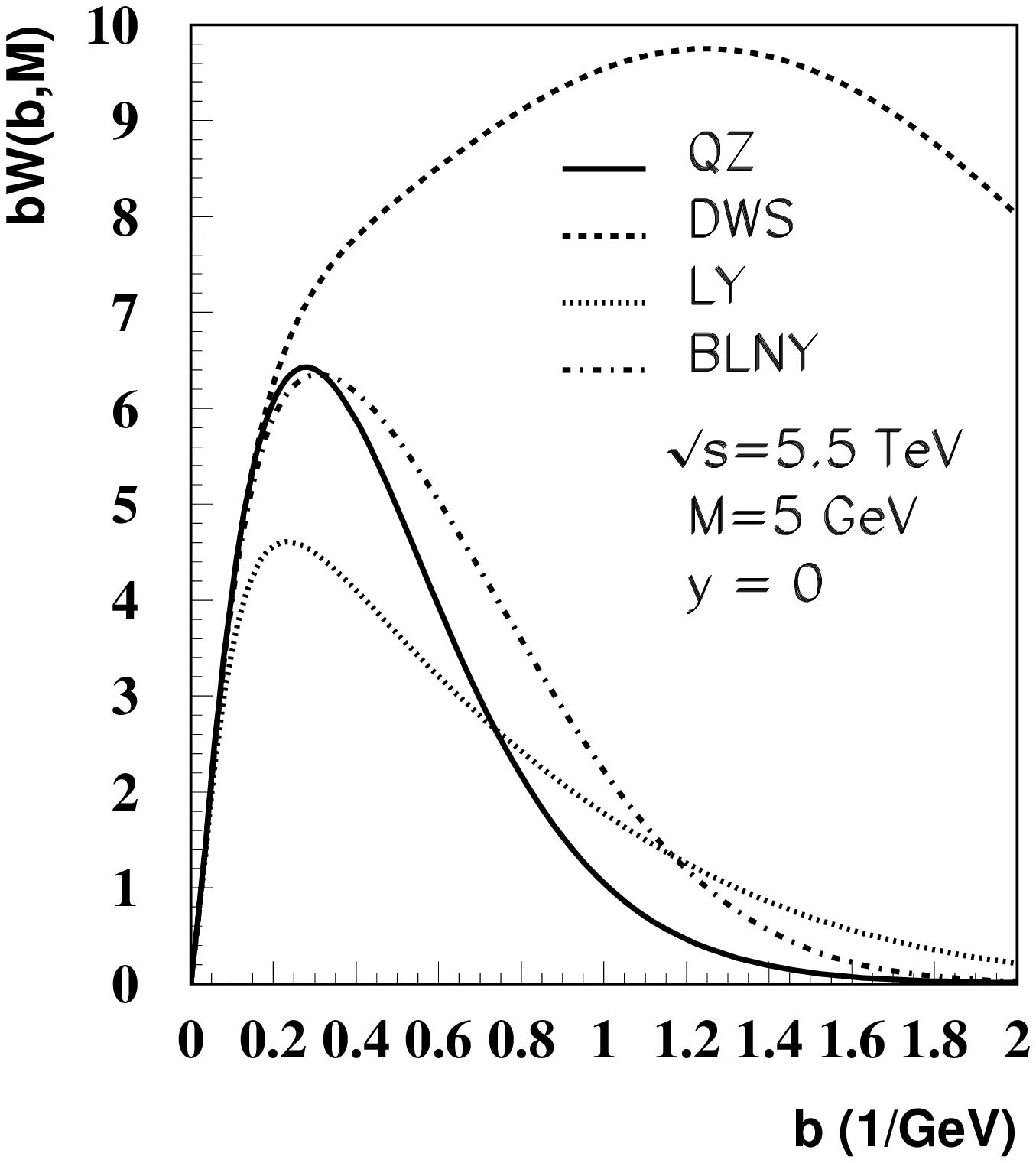}} 
\vspace{0.2in}
\caption{The $\tb$-space resummed functions $\tb W(\tb)$ in
Eq.~(\protect\ref{css-resum})
for Drell-Yan production of dilepton mass $M=5$~GeV 
at $\sqrt s=5.5$~TeV with the ZQ (solid), DWS (dashed), LY (dotted),
and BNLY (dotdashed) formalism of nonperturbative extrapolation. }
\label{fig1}
\hfill
\vspace{0.1 in}

\centerline{\includegraphics[width=6.5cm]{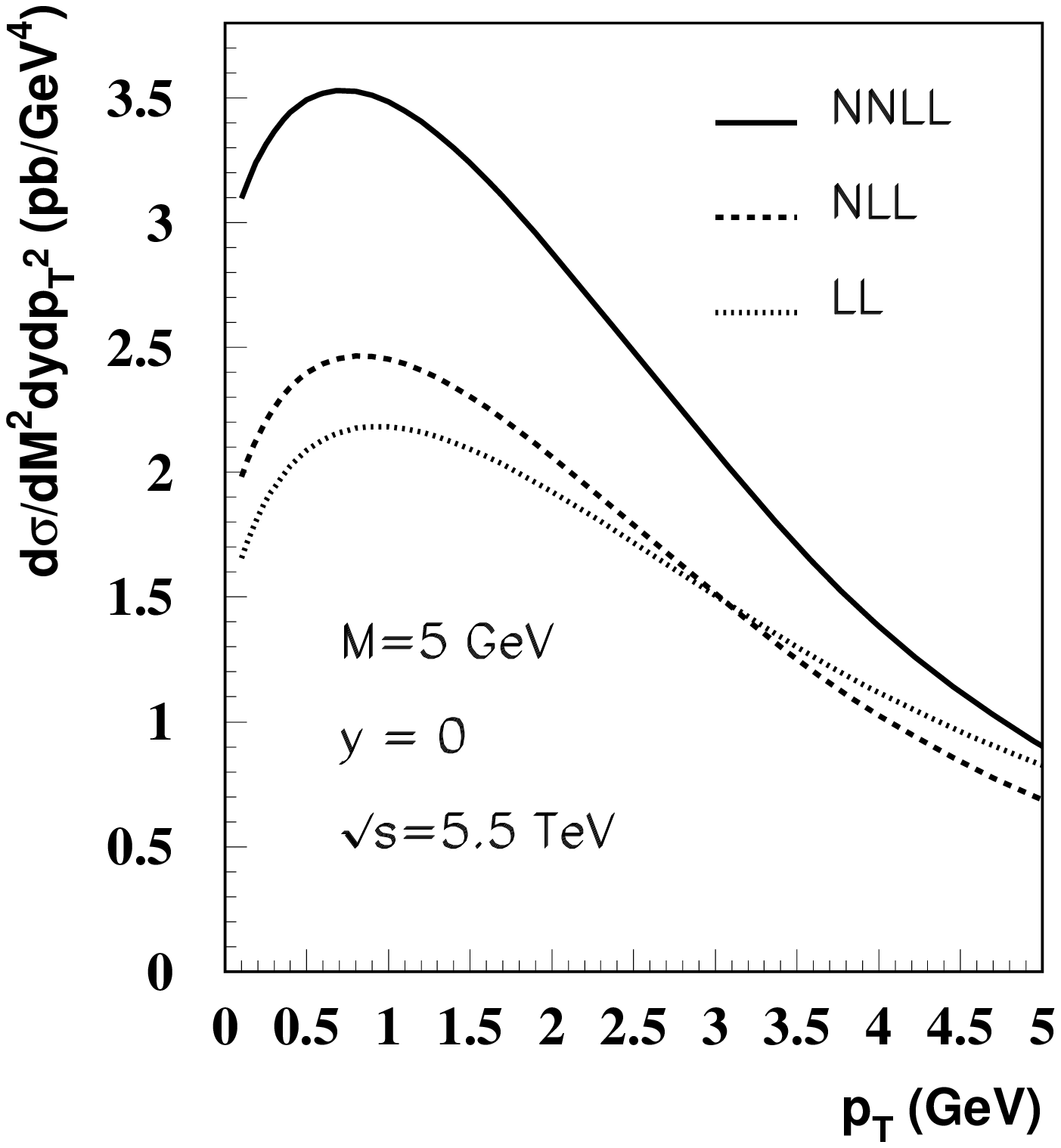}} 
\vspace{0.1in}
\caption{Differential cross section ${d\sigma /dM^2 dy dp_T^2}$ for 
production of Drell-Yan pairs of $M=5$~GeV in $pp$ collisions 
at the LHC with $y=0$, and $\sqrt{s}=5.5$~TeV with the NNLL (solid),
NLL (dashed), and LL (dotted) accuracy.}
\label{fig2}
\end{figure}

\begin{figure}
\centerline{\includegraphics[width=7.5cm]{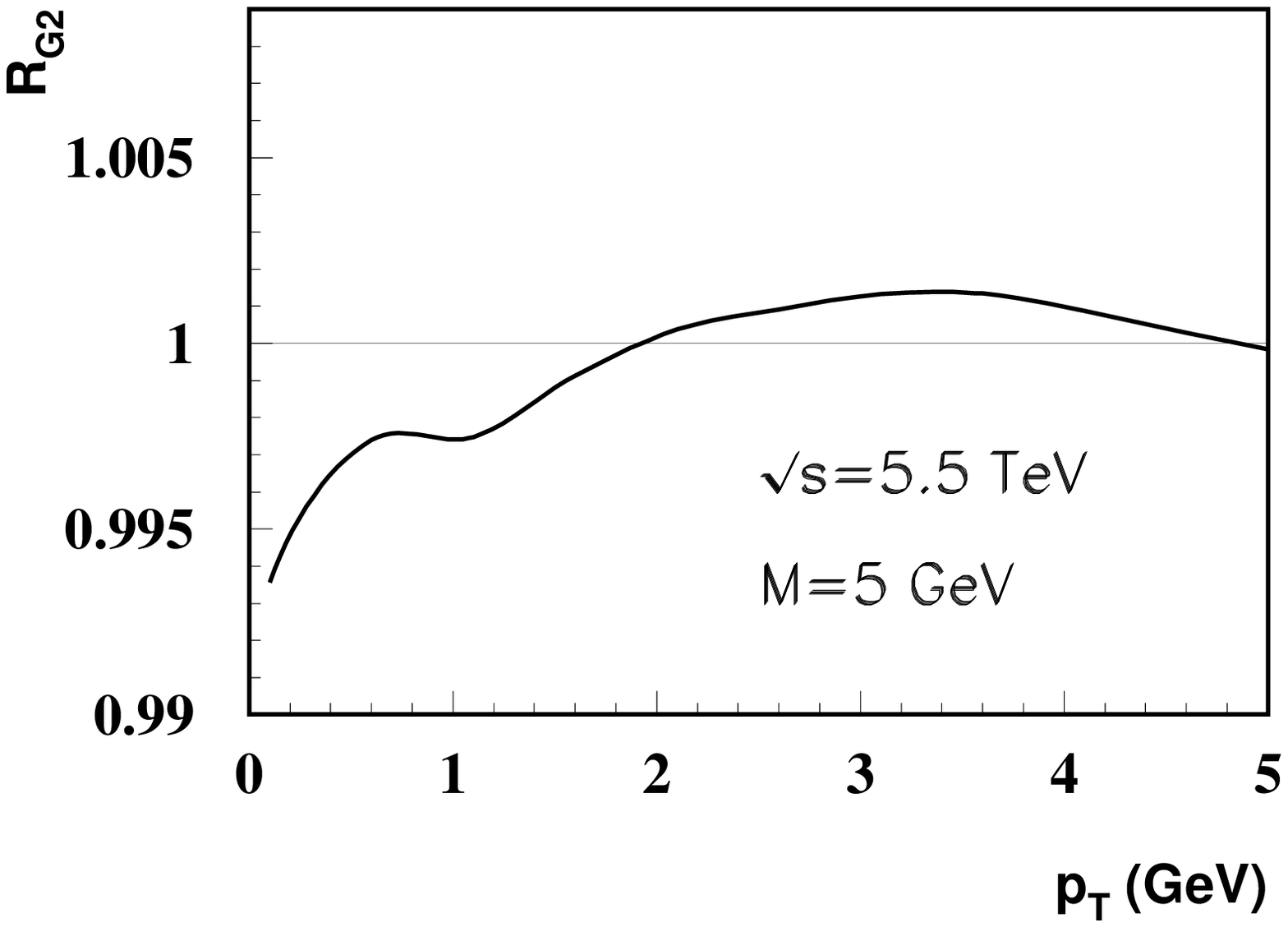}} 
\vspace{0.2in}
\caption{The ratio $R_{G_2}$ defined in Eq.~(\protect\ref{Sigma-g2})
with $G_2=$ 0.25 GeV$^2$ for the differential cross section shown in
Fig.~2.}
\label{fig3}
\hfill
\vspace{0.1 in}
\centerline{\includegraphics[width=7.5cm]{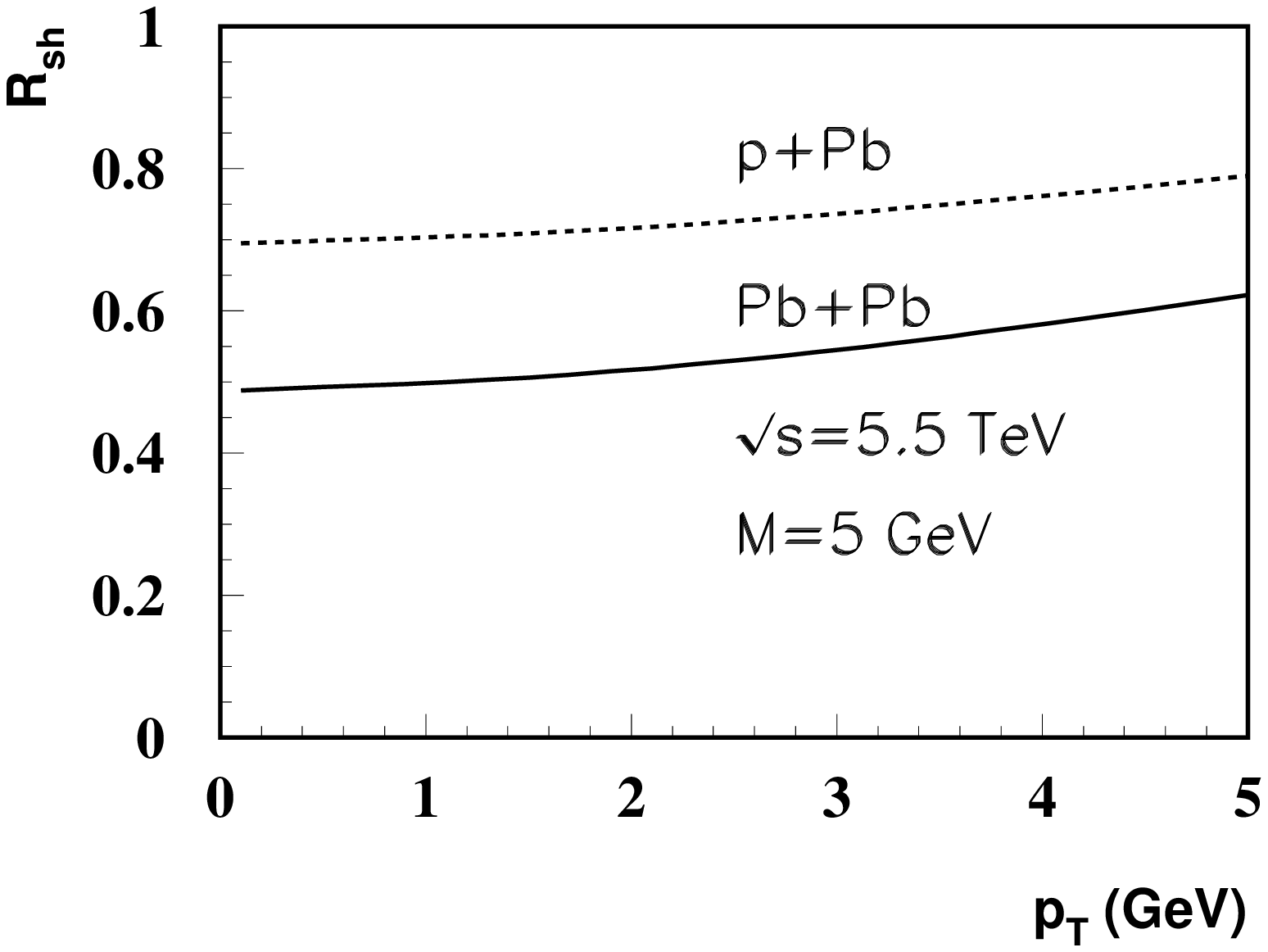}} 
\vspace{0.1in}
\caption{$R_{sh}$  as a function of $p_T$ using EKS shadowing 
in $Pb$$Pb$ collisions at $\sqrt s=5.5$ TeV.}
\label{fig4}
\end{figure}

\begin{figure}
\centerline{\includegraphics[width=6.0cm]{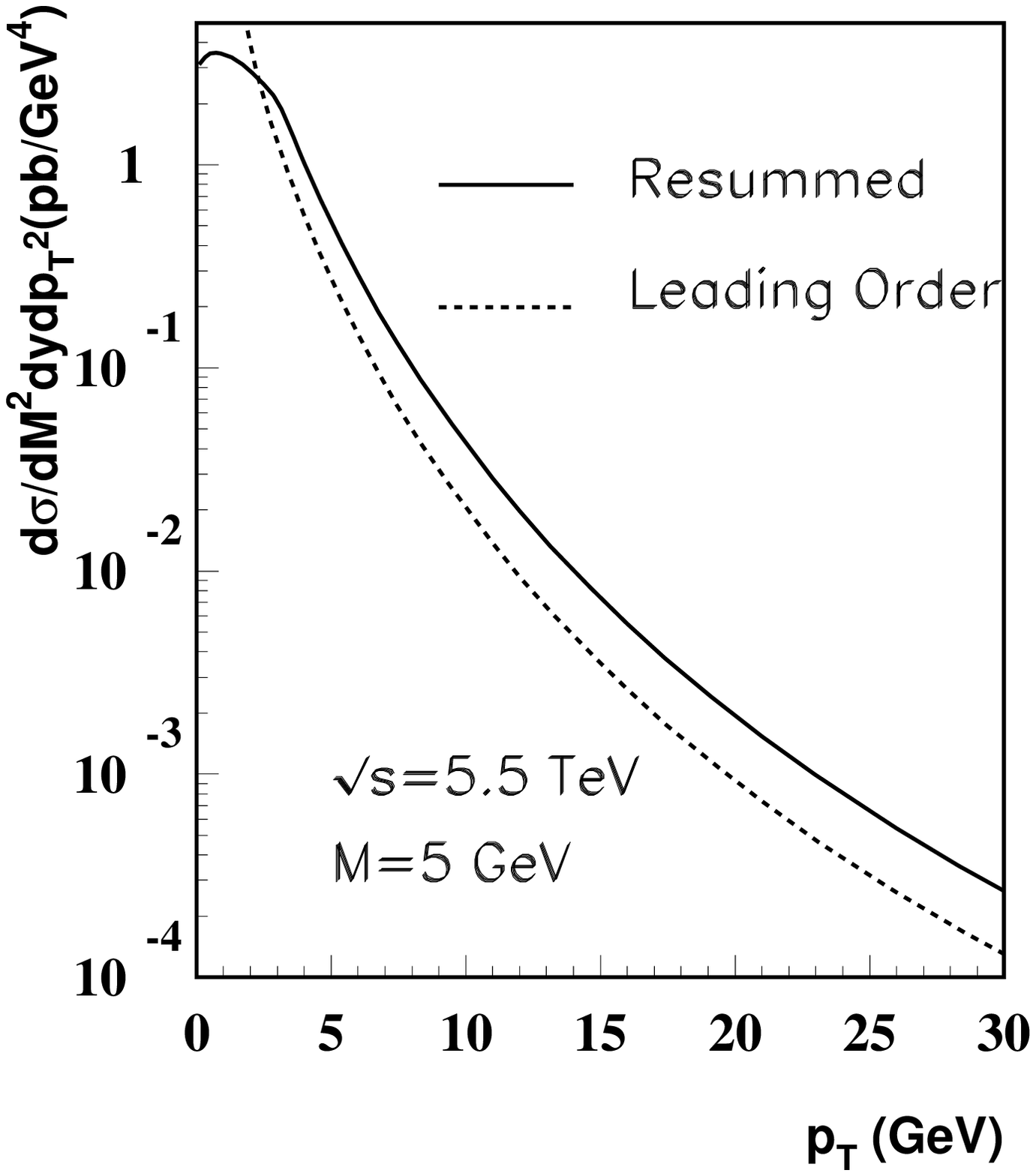}} 
\vspace{0.1in}
\caption{Differential cross section ${d\sigma / dM^2 dy dp_T^2}$ 
for production of Drell-Yan pairs of $M=5$~GeV in $pp$ collisions 
at $\sqrt s=5.5$~TeV with low and high $p_T$ resummation (solid),
in comparison to  conventional lowest result (dashed).  } 
\label{fig5}
\hfill
\vspace{0.1 in}
\centerline{\includegraphics[width=6.0cm]{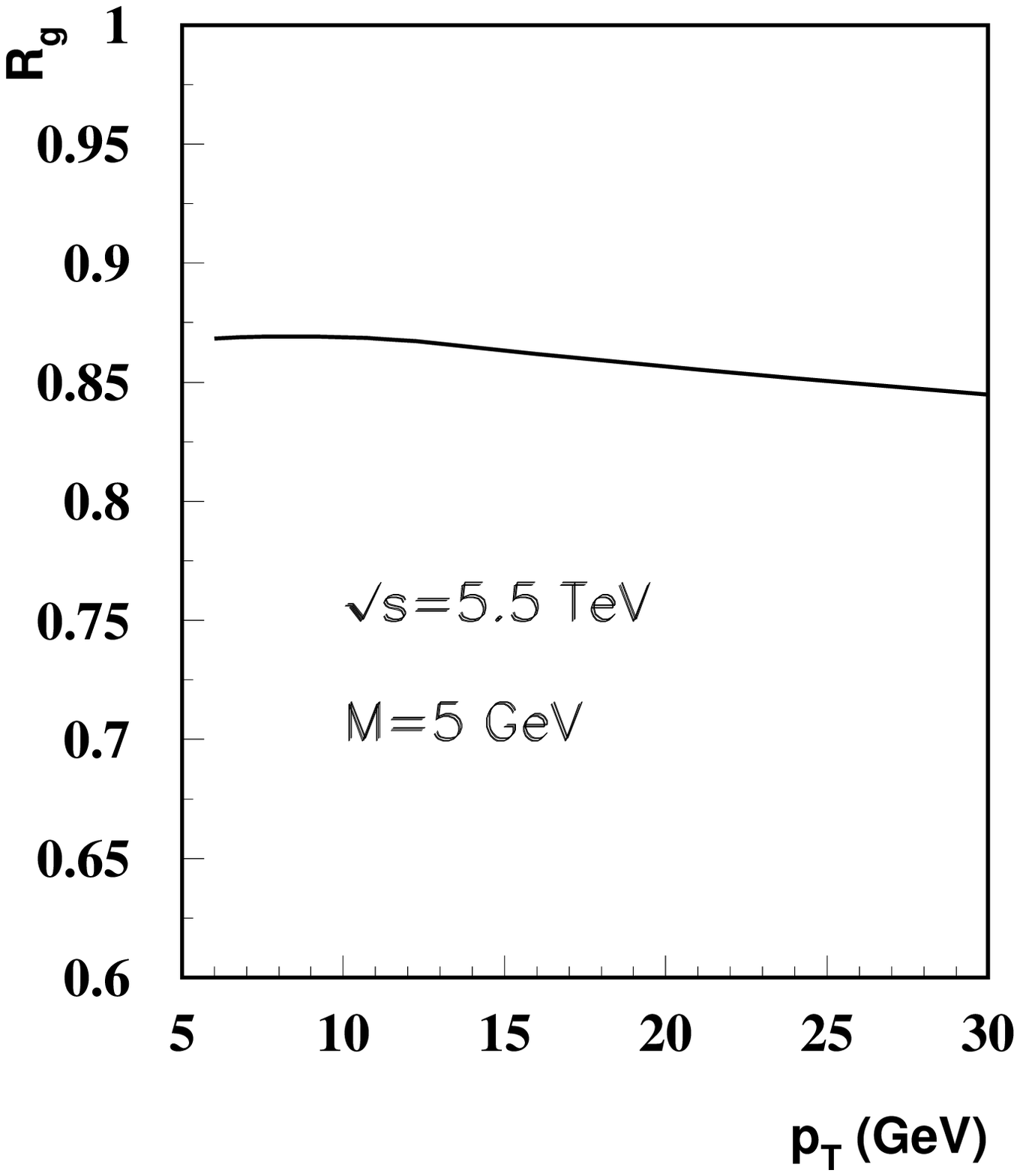}} 
\vspace{0.1in}
\caption{Ratio of gluonic over total contributions to Drell-Yan
production at the LHC, $R_{g}$, defined in 
Eq.~(\protect\ref{Rg}) with $M=5$~GeV at $\sqrt s=5.5$~TeV.} 
\label{fig6}
\end{figure}

\end{document}